\input{aipcheck}

\documentclass[
    ,final
  ]
  {aipproc}

\layoutstyle{6x9}

\begin{document}

\title{Noisy Jets and Emission Components in Galactic X-Ray Binaries}


\author{T. Belloni}{
  address={INAF - Osservatorio Astronomico di Brera, Via E. Bianchi 46, I-23807,
  			Merate, Italy}
}

\begin{abstract}
Our knowledge of the phenomenology of accretion onto
black holes has increased considerably thanks to ten years of 
observations with the RXTE satellite. 
However, only recently it has been possible
to derive a scheme for the outburst evolution of transient systems on the basis of
their spectral and timing properties, reaching a comprehensive definition 
of source states. These states are in turn linked to the ejection of
relativistic jets as observed in the radio band. Here I concentrate on
some specific aspects of this classification, 
concentrating on the  properties of the aperiodic variability
and on their link with jet ejection.
\end{abstract}

\maketitle

\section{Introduction}
\label{sec:1}

The picture of
high-energy emission from Black-Hole Transients (BHT) that is
emerging after more than a decade of observations with the RossiXTE
satellite is complex and difficult to interpret, yet it is a considerable
step forward in our knowledge. Only recently, it has been possible to 
find a sufficiently stable pattern in the spectral and timing properties
to obtain a coherent picture.  The classification of states and
state-transition derived from a few objects is applicable to most systems,
indicating that the common properties are more than what previously
known (see \cite{hombel05,bel06,hombel06} but also \cite{toms05,toms06}).
At the same time, a clear connection between X-ray and
radio properties has been found (see \cite{fenderbook,fbg04}). 

In this paper, I examine a few aspects of this picture, highlighting the 
connection between noise components, spectral properties at energies $>$20 keV, source states and radio/jet properties.

\section{Source states}
\label{sec:2}

The  color and timing analysis of GX 339-4, extended to other bright transients,
has led to a definition of {\it four} source states defined in terms of transitions between them.
For a complete definition of the four bright source states see e.g.
\cite{hombel05,bel06,bel339}, while \cite{mcrem06} adopt a different state classification based
on the parameters obtained from spectral and variability components. 
To follow the evolution of an outburst through different states, three direct observables
proved to be fundamental.
The first is the X-ray intensity of the source, expressed in count rate; the more physical 
source-integrated flux can be used, but its derivation is model dependent. The second is a measure 
of the hardness of the spectrum, namely the ratio between count rates in a hard and a soft band. 
This has the advantage of yielding a model-independent measure of the hardness of the source
spectrum: once associated to a chosen spectral decomposition, this quantity can be turned into
a physical parameter. Finally,
the third is a measure of the variability in the X-ray band, expressed as fractional rms integrated
in a reasonably wide frequency range. Of course, energy spectra and power spectra, together with 
their decomposition in different models, are important for the physical understanding, but they are 
not necessary to describe the outburst and will be discussed later.
With these three quantities, two diagrams are constructed: the Hardness-Intensity diagram (HID) and
the hardness-rms diagram.

The HID of the 2002/2003 outburst
of GX 339-4 as observed by RXTE/PCA is shown in Fig. \ref{fig:1}. The regions
corresponding to the four states discussed below are marked. 
Fig. \ref{fig:2} shows the corresponding rms-hardness diagram: on the X-axis
is the same hardness than in Fig. \ref{fig:1},
while on the Y-axis is the integrated fractional rms variability in the 0.1--64 Hz band.

In the following, I concentrate on some aspects of the states that can be identified from these diagrams:

\begin{figure}
  \includegraphics[height=.55\textheight]{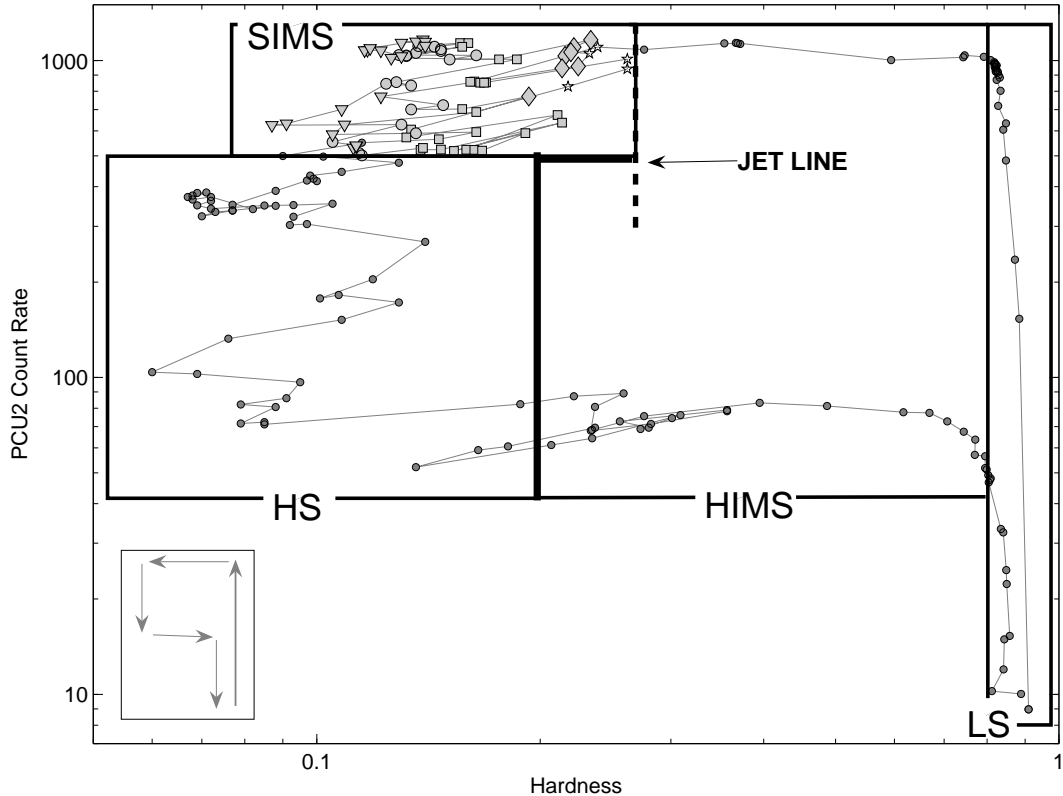}
  \caption{Hardness-Intensity diagram of the 2002-2003 outburst of GX 339-4 as observed
by the RXTE/PCA, adapted from \cite{bel06}. The lines mark the four source states described
in the text. The different symbols are described in \cite{bel339}. The dashed line shows the
position of the `jet line'. The thick line indicates the transition line that marks the presence/absence
of strong band-limited noise in the power spectra. The inset on the lower left shows the general
time evolution of the outburst along the q-shaped pattern.}
  
\label{fig:1} 
\end{figure}

\begin{itemize}

\item Two states, the Low/Hard (LS) and Hard-Intermediate (HIMS), found at the right 
of the thick vertical line in Fig. \ref{fig:1},  have a number
of aspects in common. A strong hard component is visible in their energy spectrum, usually
attributed to (thermal) Comptonization. In the LS, this can be approximated with a power law
with spectral index 1.6-1.8 and a high-energy cutoff around 100 keV. Since the source points
are distributed along an almost vertical line, it is clear that the spectral changes in this state are 
small (at least below $\sim$20 keV), while there is a flux increase by a factor of  $\sim$100.
In the HIMS, the slope is 
higher (2.0-2.4) and the high-energy cutoff moves to lower energies (see \cite{belint}).
In addition, a soft thermal disk component appears. The strong hardness variation along the
horizontal line are a combination of the steepening of the hard component and the appearance
of the thermal disk.
In both states, the power spectrum is dominated by a band-limited component, 
with the presence of a type-C
QPO in the HIMS (see \cite{casella}). The fractional rms decreases smoothly from 50 \% at the
beginning of the LS to 10\% at the end of the HIMS (see Fig. \ref{fig:2}).
These states correspond to detectable radio emission from the core source, associated with
compact jet ejection. Just before the transition to the Soft-Intermediate state, 
it was suggested that the jet velocity 
increases rapidly, giving origin to the fast relativistic jets \cite{fbg04}.

\item The other two states, the High/Soft (HS) and Soft-Intermediate (SIMS), are markedly 
different, In both, the energy spectrum is dominated by a thermal disk component, with only a weak
steep hard component visible without a high-energy cutoff (see \cite{grove}). Again, the small hardness
variations in this state (notice that the hardness scale in Fig. \ref{fig:1} is logarithmic) indicates
that there are no strong spectral changes in the HS. The power spectrum 
does not show a strong band-limited noise, but rather a weaker power-law component. In the
SIMS, strong type-A/B QPOs are observed, while in the HS occasionally weak QPOs are observed,
which cannot be easily identified with the ABC types (see \cite{casella,bel339}).
As one can see from Fig. \ref{fig:2}, the integrated fractional rms is below 10\% and goes down 
as low as 1\% at  the soft end. Moreover, the HS points (triangles and dots) follow a continuation
of the rms-hardness correlation of the LS/HIMS, while the SIMS points deviate from it, being
at lower rms values (boxed area in Fig. \ref{fig:2}). 
Also notice that when the source returned to the HIMS at the end of the 
outburst (black circles), only one point can be attributed to a SIMS, while the others
followed the rms-hardness correlation all the way to the right.
These states, to the left of the `jet line', correspond to non-detectable emission from the central
source (see \cite{corbel,fbg04}). However, the multi-wavelength analysis of an outburst of three 
transients showed that the core radio emission turns on only when the source reaches the LS
(\cite{homanir,kalemci,kalemci06}).

\end{itemize}

\begin{figure}
  \includegraphics[height=.55\textheight]{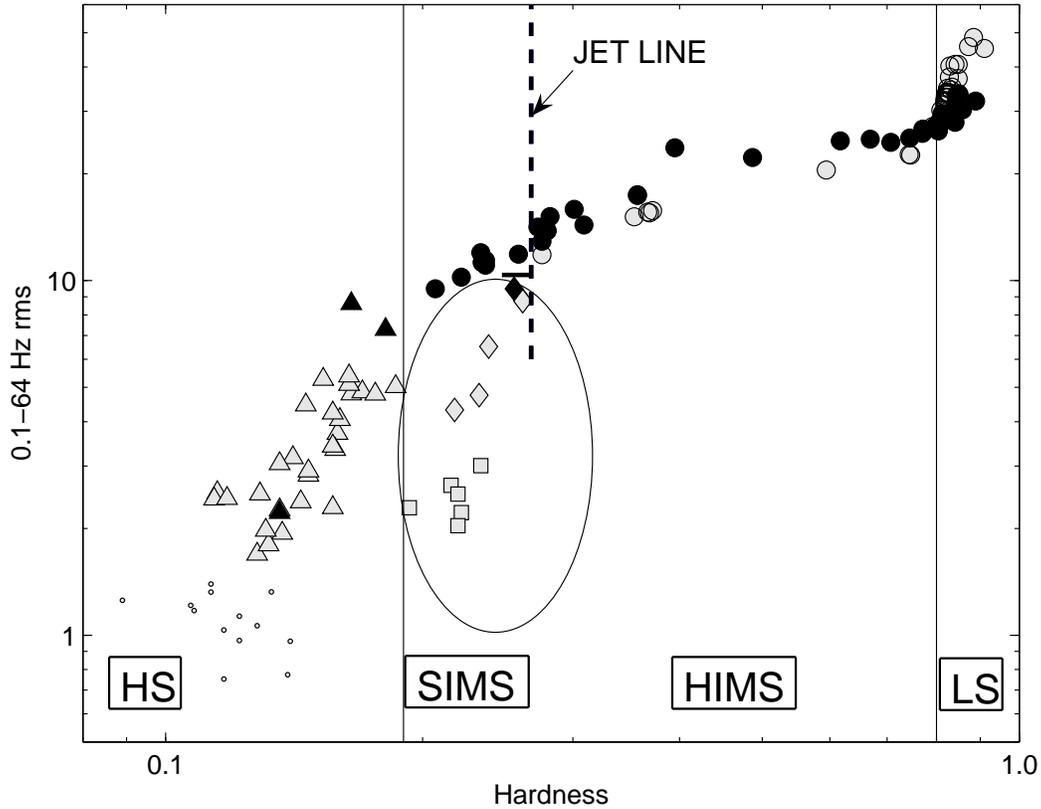}
  \caption{Rms-hardness diagram corresponding to the observations in Fig. \ref{fig:1}. The hardness
intervals corresponding to the four states are marked. Gray points correspond to the first part of the
outburst (followed right to left), black points to the second part (left to right). Different symbols 
indicate different power spectral shapes. Circles: strong band-limited noise (and type-C QPO);
diamonds: type-B QPO; squares: type-A QPO; triangles: unidentified QPO; dots: only weak signal.
The ellipse shows the SIMS points, characterized by a drop in rms.}
\label{fig:2} 
\end{figure}

Obviously, the four states can be grouped into two `classes': the hard and radio-loud ones
(LS and HIMS, although radio emission is observed in the HIMS only at high luminosities, see
\cite{kalemci,kalemci06}) and the soft and radio-quiet ones (HS and SIMS). The evolution from hard
to soft causes a crossing of the jet line: the velocity of the outflow increases rapidly when approaching
the line, creating fast propagating shocks in the jet \cite{fbg04}. The evolution from soft to hard
simply leads to the (slow) formation of a new outflow, only mildly relativistic, and cannot give rise
to fast relativistic ejections.

\begin{figure}
  \includegraphics[height=.30\textheight]{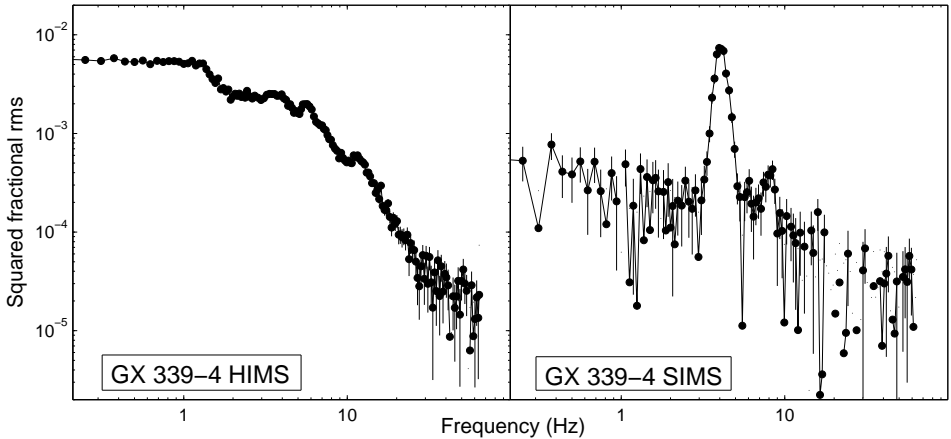}
  \caption{Power density spectra of two RXTE observations of GX 339-4 from 
  2004 August 15. The two observations were separated by about 8.5 hours. 
  Left panel: HIMS before the transition; right panel: SIMS after the transition. For
  a full analysis of these data, see \cite{belint}.
  }
\label{fig:3} 
\end{figure}

The focus here is obviously on the transitions between states, which mark very clearly the 
division between them. The transition between LS and the HIMS is {\it clearly} 
marked by changes in the infrared/X-ray correlation and can be placed very precisely in the
evolution of the outburst \cite{homanir,bel339}. The transition between HIMS and SIMS
is marked by abrupt changes in the timing properties, with the appearing of type-A/B QPOs
and the dropping of the noise level (see Fig.  \ref{fig:3}). 
From SIMS to HS, a recovery of the noise level and disappearing of type-A/B QPOs is observed.

\section{Time variability as a tracer}
\label{sec:3}

No light curve is shown in the figures here. It is not necessary to examine the time evolution,
as the evolution of the system can be synthesized with only the two plots in Figs. \ref{fig:1} and
\ref{fig:2}. In particular, Fig. \ref{fig:2} is particularly important, as it shows at the same time 
the color (spectral) and rms (timing) evolution. The presence/absence of specific noise components
and/or QPOs, with related problems of significance of detection, is not necessary in order to 
characterize the source states in this diagrams.

In Fig. \ref{fig:1}, the source follows a very different path moving from hard to soft at the beginning
of the outburst than it does the opposite transition at the end of the outburst. This hysteresis cycle has
been discussed in the past (see \cite{miya,macca}). However, notice that there is no trace of
hysteresis in Fig.  \ref{fig:2}: here the source traces the same path throughout the early and 
late phases of the outburst. Since the bright and weak transitions take place at different values
of mass accretion rate, this indicates that the global timing properties do not depend on this
parameter.

It is evident from Fig. \ref{fig:2} that the timing
properties are tracers of the evolution of the accretion flow and the jet properties. The crossing
of the jet line can be identified by the drop in integrated rms 
and by the appearance
of type-A/B QPOs, easily identifiable in the power spectra. 
Notice that the recent analysis of the radio/X-ray properties of a number of systems
suggest that  the association between 
crossing the jet line and entering the SIMS is not precise: the ejection time of the jet 
can lead or lag the start of SIMS by a few days (\cite{fennew}).
 
 In addition to the integrated variability properties, specific features in the power
 spectra are important players in the game. 
Type-C QPOs and band-limited noise
components provide characteristic frequencies that trace the evolution of the
accretion during the early and final part of the outburst; in the hard states,
the strong correlations between hard X-ray emission and radio flux led to models
for the jet production and X-ray emission which need to take these timing
tracers into account (see e.g. \cite{meier}). Type-A/B QPOs are much less studied
and their origin is unclear. However, the transient presence of type-B QPOs
could be important for the study of the physical conditions of the accretion
flow as the source crosses the jet line. The small range of long-term
variability of their centroid frequency, its fast short-term variability (see
\cite{nespoli,bel339}) and the absence of band-limited noise
components are key ingredients for their understanding (see \cite{casella}).

\section{Conclusions: noisy accretion and ejection}
\label{sec:4}

The general picture that is emerging is becoming more and more clear and it can be 
applied to a large sample of systems \cite{hombel06}. In particular, the details of the
changes in the X-ray emission when crossing the jet line appear crucial for the
modeling of the accretion/ejection connection. Although the timing properties
change on a very short time scale, at low ($<$20 keV) energies spectral changes are minimal,
as can be seen by the small difference in X-ray hardness (see e.g. \cite{nespoli}).
Recently, evidence for a possible
fast change in the spectral properties at energies $>$20 keV was gathered \cite{belint},
although more high-energy data on this elusive transition should be obtained.

Although timing analysis of the fast variability of BHTs can
give us direct measurements of important parameters of the accretion flow,
up to now we do not have unique models that permit this. Recent results show
that
a clear association can be made between type-C QPO, strong band-limited noise
and the presence of a relativistic jet. In the framework of unifying models,
these results could play an important role.

\bibliographystyle{aipproc}
\bibliography{sample}

\IfFileExists{\jobname.bbl}{}
 {\typeout{}
  \typeout{******************************************}
  \typeout{** Please run "bibtex \jobname" to obtain}
  \typeout{** the bibliography and then re-run LaTeX}
  \typeout{** twice to fix the references!}
  \typeout{******************************************}
  \typeout{}
 }

%
%

%
%

\end{document}